\documentclass[aps,prl,twocolumn,superscriptaddress]{revtex4}
\usepackage[dvips]{graphicx}

\begin{document}


\title{Nanoscale electromechanical resonators as probes of the charge density wave transition in NbSe$_2$}


\author{Shamashis~Sengupta}
\author{Hari~S.~Solanki}
\author{Vibhor~Singh}
\author{Sajal~Dhara}
\affiliation{Department of Condensed Matter Physics and Materials Science, Tata Institute of Fundamental Research, Homi Bhabha Road,
Mumbai 400005, India}
\author{Mandar~M.~Deshmukh}
\email[]{deshmukh@tifr.res.in}
\affiliation{Department of Condensed Matter Physics and Materials Science, Tata Institute of Fundamental Research, Homi Bhabha Road,
Mumbai 400005, India}


\date{\today}

\begin{abstract}
We study nanoelectromechanical resonators fabricated from suspended flakes of NbSe$_2$ (thickness$\sim$30-50 nm) to probe charge density wave (CDW) physics at nanoscale. Variation of elastic and electronic properties accompanying the CDW phase transition (around 30 K) are investigated simultaneously using the devices as self-sensing heterodyne mixers.  Elastic modulus is observed to change by 10 per cent, an amount significantly larger than what had been reported earlier in the case of bulk crystals. We also study the modulation of conductance by electric field effect, and examine its relation to the order parameter and the CDW energy gap at the Fermi surface.
\end{abstract}


\maketitle



Charge density waves (CDW) \cite{gruner} result from the coupling between electrons and phonons, which distorts the electron distribution inside the crystal and the lattice also undergoes a periodic deformation. The experimental techniques used so far to study charge density waves include neutron scattering \cite{neutron}, electrical transport measurements \cite{transport},
vibrating reed technique \cite{barmatz}, angle-resolved photoemission spectroscopy (ARPES) \cite{ARPES} amongst others. The CDW modifies both the elastic and electrical transport properties of the system. Many questions regarding the microscopic origin of CDW in transition metal dichalcogenides (TMD), which are quasi-2D materials, have still not been resolved. In the last few years, new insights have emerged about the formation of CDW in TMD (issues like Dirac fermion excitations \cite{Castro Neto} and feasibility of Peierls instability \cite{doubt}). In this letter, we will focus on the CDW transition in 2H-NbSe$_2$ (a TMD), which has elicited a lot of interest for its unique properties and is known to support an incommensurate CDW at low temperatures (below 35 K).

With technological advances in small scale device fabrication, CDW in mesoscale and nanoscale systems \cite{slot, calandra} have become the subject of experimental and theoretical studies. The isolation of atomically thin NbSe$_2$ layers and transport measurements \cite{frindt} on them have been reported. Our study has been conducted on nanoelectromechanical resonators made from suspended flakes of NbSe$_2$ (thickness$\sim$30-50 nm, i.e., 45-80 monolayers) in doubly clamped geometry. Resonators based on nanoelectromechanical systems (NEMS) have been realized earlier in carbon nanotubes \cite{sazonova paper}, nanowires \cite{roukes, hari} and graphene \cite{vibhor}, to name a few materials. They have been used in mass sensing, studying the spin-torque effect and Casimir force etc. (Ref. \cite{hari} and others cited therein) Using them to investigate the CDW transition in NbSe$_2$, we demonstrate here that such resonators can also act as highly sensitive probes for concomitant structural and electronic phase transitions at nanoscale.

\begin{figure}
\includegraphics[width=80mm]{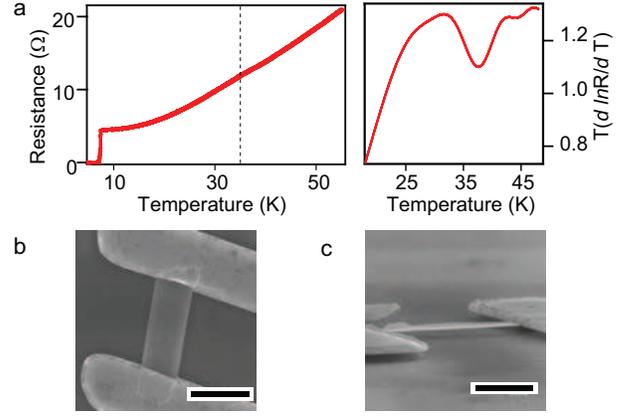}
\caption{\label{fig:figure1} (a) (Left) Resistance ($R$) as a function of temperature ($T$) for an on-substrate NbSe$_2$ device. There is a feeble and broad peak around 35 K (marked by the dashed line), characteristic of the CDW transition. (Right) $T\frac{d(ln R)}{dT}$ plot showing the CDW feature prominently. (b) Scanning electron microscope (SEM) image of a suspended NbSe$_2$ flake. Scale bar: 2$\mu$m. (c) SEM image showing side view of a different suspended flake. Scale bar: 1$\mu$m.}
\end{figure}

To fabricate our devices, we deposit NbSe$_2$ flakes on a silicon wafer (coated with 300 nm of insulating SiO$_2$) by mechanical exfoliation \cite{frindt}. Standard electron beam lithography techniques, followed by sputtering Cr and Au, are used to establish metallic contacts to these flakes. Before metallization of electrodes, the sample is cleaned in Ar plasma to remove oxide and organic residue. DC electrical transport measurements on such devices show a feeble and broad peak in resistance around 35 K due to CDW transition (Fig. 1(a)). They undergo superconducting transition at 7.2 K. These values are similar to observations in bulk crystals.

The contacted NbSe$_2$ flakes are suspended by etching away the SiO$_2$ underneath in a dilute buffered HF solution. The typical trench depth is 180 nm. Fig. 1(b) and 1(c) show scanning electron microscope (SEM) images of our suspended devices (typical length $\sim$ 2-3 $\mu$m). To actuate vibrational motion in NbSe$_2$ devices, they are used as heterodyne mixers (Fig. 2(a)). The backplane of the silicon wafer serves as the gate electrode. An RF voltage (amplitude denoted by $\widetilde{V}_g^{ac}$) at frequency $f$ and a DC bias ($V_g^{dc}$) are simultaneously applied to the gate with a bias-tee ($V_g^{dc} \gg \widetilde{V}_g^{ac}$). Due to its capacitive coupling to the gate, the NbSe$_2$ flake experiences a vertical driving force at the RF frequency $f$ \cite{force}. The source-drain voltage (amplitude $\widetilde{V}_{sd}$) is applied at a frequency $f$+$\Delta$$f$, shifted by $\Delta$$f$ with respect to the RF gate signal. $\Delta$$f$ ($\sim$6-20 kHz) is kept constant throughout the measurement. The current through the device at the difference frequency $\Delta$$f$ is called the mixing current ($I_{mix}$) \cite{thesis sazonova, hari, knobel}. \begin{equation}
I_{mix}=\frac{dG}{dq}(A\xi_f + B)
\end{equation} where $G$ is the conductance of the NbSe$_2$ flake, $q$ is the charge induced by $V_g^{dc}$ and $\xi_f$ is the amplitude of mechanical vibration (at applied frequency $f$) along the $z$-direction perpendicular to the plane of the suspended flake (see Fig. 2(a)). $A$ and $B$ are factors that depend upon applied voltages and the device geometry \cite{full equation}.

Since the measurable mixing current depends upon how well the conductance is modulated by inducing charges (the factor $\frac{dG}{dq}$ in Eqn. 1), this technique of heterodyne mixing had so far been applied to materials which show a considerable gating effect (change in conductance as a function of gate voltage), like carbon nanotubes and semiconductor nanowires. Although NbSe$_2$ flakes having thickness of tens of nanometers are metallic, this method works when the resonator is excited in the linear regime to such a degree that the resonance contribution to $I_{mix}$ is distinguishable against the noise floor \cite{thorne}. Fig. 2(b) shows a plot of mixing current versus frequency for Device 1 at 10 K temperature. The peak at 23.9 MHz reflects the mechanical resonance. Quality factor of resonance, on fitting to Eqn. 1 (see also Ref. \cite{full equation}), is 215. In practice, the $\Delta f$ signal measured by us is $I_{mix}$ (given in Eqn. 1) plus a background (of the order of hundred pA) arising out of other sources in the circuit.

\begin{figure}
\includegraphics[width=80mm]{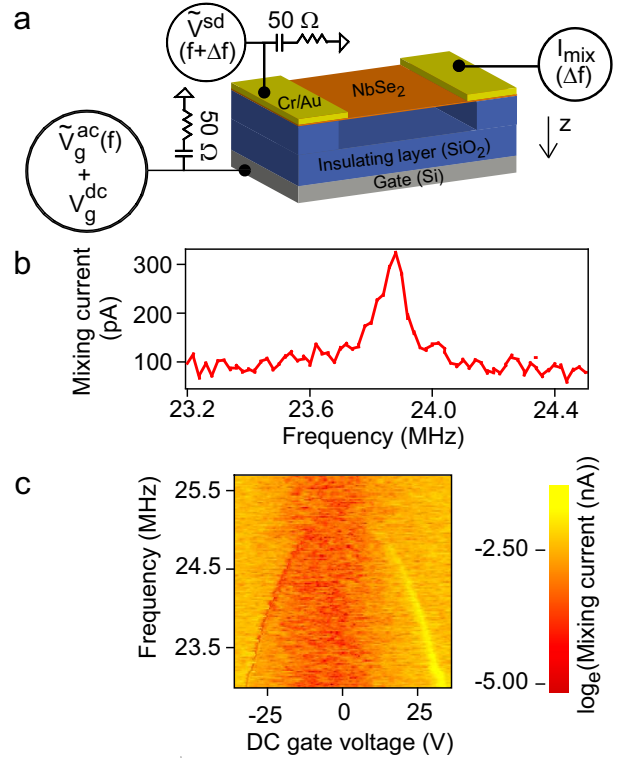}
\caption{\label{fig:figure2} (a) Schematic of actuation technique. (b) Measurement at 10 K, showing mixing current ($I_{mix}$)as a function of driving frequency $f$ for Device 1. ($V_g^{dc}$=32 V, $\Delta f$=12 kHz). (c) Colourscale plot of $I_{mix}$ as a function of both frequency and DC gate voltage $V_g^{dc}$ for Device 1 at 10 K. The points corresponding to mechanical resonance of the device trace out a parabolic curve. ($\Delta f$=12 kHz).
}
\end{figure}

At room temperature, when the stress in the flake is negligible \cite{sapmaz}, the natural frequency is given by $f_0$=3.56$\sqrt{\frac{EI}{\rho S}}\frac{1}{L^2}$. Here $L$ is the length of the suspended flake, $E$ the elastic modulus, $I$ the inertia moment (=$\frac{1}{12}$(width)$\times$(thickness)$^3$), $\rho$ the mass density (=6.467 g/cm$^3$) \cite{nbse2 density} and $S$ the cross-sectional area. From room temperature resonant frequency measurements, $E$ of most of these NbSe$_2$ flakes of approximately rectangular shape are calculated to lie in the range $1.0 - 1.6\times10^{11}$ Pa. On cooling them down, the metallic electrodes and NbSe$_2$ contract, enhancing the tension $\Gamma$ and strain $\varepsilon$ in the resonator. In the high tension regime \cite{sapmaz},
\begin{equation}
f_0 = \sqrt{\frac{EI}{\rho S}}(\frac{1}{2L}\sqrt{\varepsilon \frac{S}{I}} + \frac{1}{L^2})
\end{equation}
In Device 1 (length 3.3$\mu$m, width 1.7$\mu$m, thickness 50 nm), $f_0$ changed from 18.6 MHz at room temperature to 25.2 MHz at 10 K. Its elastic modulus at room temperature is $1.0\times10^{11}$ Pa. Assuming $E$ remains the same even at low temperatures and using Eqn. 2, the stress ($=\Gamma/S$) at 10 K is calculated to be $1.1\times10^8$ Pa ($\sim$ 1 kBar) and the strain in the device is $1.1\times10^{-3}$.

Fig. 2(c) shows a colourscale plot of mixing current as a function of frequency
and $V_g^{dc}$ (at 10 K). The points where mixing current amplitude changes sharply (corresponding to mechanical resonance of the device) are distinguishable against the background as the parabola-shaped contour. The resonant frequency decreases away from 0 V (on both sides) as a function of $V_g^{dc}$. This behavior \cite{negative dispersion} is typical of electrostatically actuated resonators (seen earlier in nanowire \cite{hari}, nanotube \cite{thesis sazonova} and graphene \cite{vibhor} based NEMS devices).

Temperature scans were conducted by slowly heating the devices inside a helium flow cryostat (controlled temperature sweep at typical rate $\sim$ 75 mK/min in the presence of exchange gas to ensure proper thermalization). Fig. 3(a) shows data from Device 2 (thickness 35 nm). The striking feature is the huge change in resonant frequency around the CDW transition. In the temperature range between 28 K and 40 K, change in strain due to thermal expansion of Cr/Au electrodes being negligible, the resonant frequency can be related to the elastic modulus as $f_0 \propto \sqrt{E}$ (see Eqn. 2). From the observed resonant frequencies at these two temperatures, the fractional change in $E$ across the CDW transition is estimated as 10 percent \cite{strain correction}. This signifies a huge degree of softening of acoustic phonons in the CDW state, exceeding by far observations in earlier studies on bulk samples \cite{barmatz, jericho, skolnick}. In their vibrating reed study \cite{barmatz}, Barmatz \emph{et al.} reported a 0.2 per cent change in elastic modulus $E$, and  unlike their results we do not see a negative peak in $E$ as a function of $T$ (temperature). The normal - incommensurate CDW phase transition in NbSe$_2$ is accepted to be second order. Landau theory calculations suggest a weak first order transition in the presence of impurities \cite{mcmillan}. Lattice distortions and defects can play an extremely vital role in such small samples. Impurities can strongly pin the incommensurate CDW, modifying its coupling to the lattice and this may be one of the reasons for the unexpected behavior of elastic modulus observed by us.

\begin{figure}
\includegraphics[width=80mm]{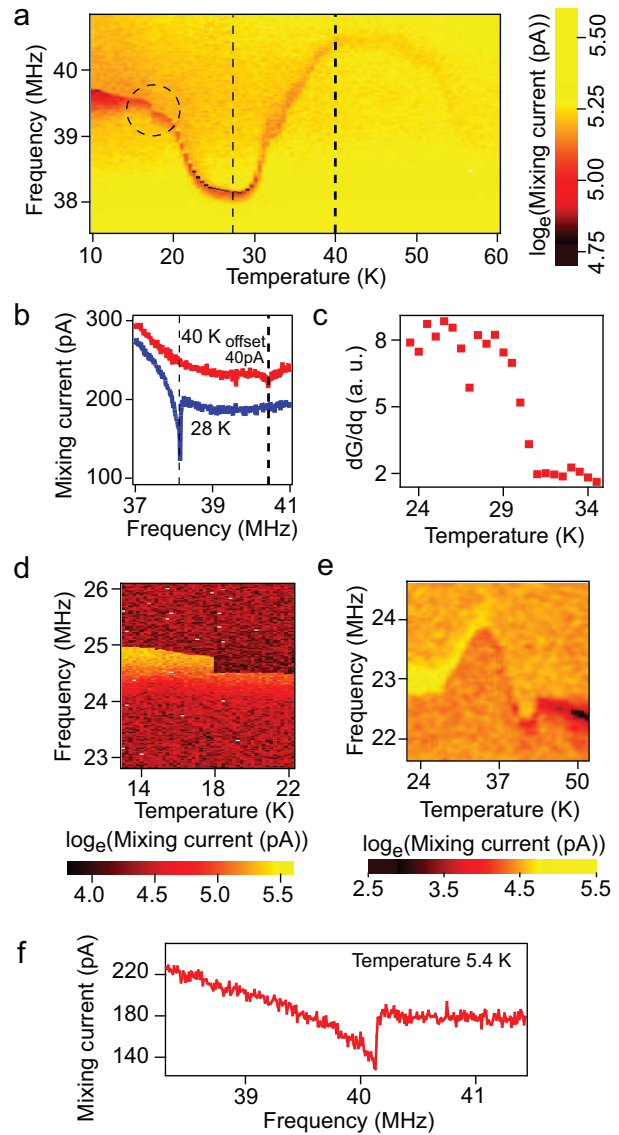}
\caption{\label{fig:figure2} (a) Colourscale plot of mixing current, $I_{mix}$, as a function of driving frequency $f$, scanned over a range of temperatures for Device 2. The resonant frequency starts to increase sharply after 30 K, a signature of transition from the charge density wave to normal state. An abrupt change (dashed circle) is also noticeable at 18 K. ($V_g^{dc} =$ 42 V, $\Delta f = 6$ kHz). (b) $I_{mix}$ vs. $f$, at two different temperatures (marked by dashed lines in Fig.3(a), below and above the CDW transition temperature. (c) Variation of $\frac{dG}{dq}$ (arbitrary units) with temperature. (d) Finer scan of $I_{mix}$ vs. $f$ and temperature around 18 K for Device 1. ($V_g^{dc} =$ 28 V, $\Delta f$ = 12 kHz). (e) $I_{mix}$ as a function of $f$ and temperature for Device 1 around the CDW transition. ($V_g^{dc} =$ - 32 V, $\Delta f = 12$ kHz). (f) Resonance measurement linescan at 5.4 K (in Device 2) when NbSe$_2$ is a superconductor. ($V_g^{dc} =$ 42 V, $\Delta f = 6$ kHz).}
\end{figure}

Next we focus on the aspects of the above-mentioned data which highlight electronic properties of the CDW. The measured resonance feature is prominent in the CDW state (below 30 K) but fades out at higher temperatures (Fig. 3(a)). This rapid change is shown by the two plots (at 28 K and 40 K) in Fig. 3(b). The sharp peak in $I_{mix}$ corresponding to mechanical resonance almost disappears just after the CDW transition. This occurs because $\frac{dG}{dq}$ reduces drastically while ramping up the temperature at the CDW transition. The height of the peak in $I_{mix}$ with respect to the background is a measure of $\frac{dG}{dq}$. Its dependence upon the CDW order parameter (energy gap at Fermi surface) is borne out clearly in Fig. 3(c). Power law fit to $\frac{dG}{dq}\sim(T_c-T)^\beta$ gives $T_c$ = 30.6 K and $\beta$ = 0.47$\pm$0.1. This estimate of $\beta$ is surprisingly close to 0.5, the critical exponent of the order parameter in Ginzburg-Landau theory. When the CDW is pinned and gate voltage $V_g^{dc}$ is applied, induced charges will modify the carrier density at the edges of the `valence' and `conduction' bands near the Fermi energy, leading to a change in the conductance $G$. (We draw an analogy here to the band gap and electrostatic gating effect in a semiconductor \cite{ibach}.) However, the energy gap disappears upon transition to the normal state. Conduction becomes metallic, sensitive only to the electron concentration at the Fermi energy. In a metal, induced charges can not alter the Fermi level significantly because of the large density of states. Therefore, $\frac{dG}{dq}$ (and also the $I_{mix}$ signal of resonance) is larger in the presence of a CDW than in the normal state. Recently, the existence of a CDW energy gap at the Fermi surface of NbSe$_2$ has been confirmed by Borisenko \emph{et al.} using ARPES \cite{ARPES}. The gap at 19.5 K is 2.4 meV, greater than $kT$ ($\sim$1.7 meV). So, the analogy to a semiconductor is justified.

CDW affects only a small fraction of the Fermi surface in NbSe$_2$. There are enough free electrons, even below CDW $T_c$, to screen out any electric field in the bulk. The modulation in conductance due to induced charges, giving rise to a measurable $I_{mix}$, takes place in a few layers at the surface of the NbSe$_2$ flake which faces the gate electrode. It is interesting to note that we can detect the resonance feature prominently even when NbSe$_2$ is superconducting (Fig. 3(f)). $\frac{dG}{dq}$ does not show any change at or near 7 K, suggesting that the CDW part of the Fermi surface is unaffected by superconductivity (a fact corroborated by ARPES data \cite{ARPES}) and the CDW energy gap exists even in the superconducting state (at least at the surface).

At 18 K, an abrupt decrease in resonant frequency is observed for both Devices 1 and 2 (Fig. 3(d) and 3(a)). In Fig. 3(d) (data from Device 1), the sharp drop of 280 kHz corresponds to a 2.1 per cent reduction in $E$. (In Device 2, the change in $E$ at 18 K is 0.7 per cent.) To the best of our knowledge, no anomaly has been reported earlier in the elastic modulus or thermal expansion coefficient of bulk NbSe$_2$ crystals around this temperature. Apart from lattice imperfections and impurities, the large stress ($\sim$1 kBar) in our samples can possibly be an important factor in determining the elastic properties. Following the article on structural phase transitions by Rehwald \cite{Rehwald}, the behavior of $E$ close to 18 K suggests a second order phase transition, where the strain couples quadratically to the order parameter in the Landau free energy. Further experiments are essential.

Fig. 3(e) shows the variation of $E$ near CDW transition of Device 1. However, on a few experimental runs on this device, we have observed larger divergent behaviour. Occasionally, we have also seen the absence of prominent CDW features. These may be signatures of metastability or near-critical-point behaviour. This is not well understood and more detailed studies are necessary to probe this phenomenon.

In conclusion, we have demonstrated a new way to look into CDW physics at nanoscale which, as our results show, can be different and highly unexpected even for well-known materials. This technique can query both elastic and electronic properties, and has immense potential for a wider range of applications to strongly correlated phenomena, including systems undergoing structural phase transitions and non-equilibrium properties of few vortex dynamics in nanoscale superconductors.

We would like to thank B. Parkinson for the crystals of NbSe$_2$, and acknowledge helpful discussions with A. Allain, S. Bhattacharya, J. Brill, M. Calandra, I. Mazin, S. Ramakrishnan, K. Ramola, and V. Tripathi. This work was supported by the Government of India.



\begin{thebibliography}{99}
\bibitem{force}
The amplitude of driving force experienced by the resonator is $\frac{dC_g}{dz}V_g^{dc}\widetilde{V}_g^{ac}$, where $C_g$ is the gate capacitance.
\bibitem{negative dispersion}
There are two competing factors involved here: (i) The applied DC electric field tends to soften the stiffness of the resonator (by a factor proportional to $d^2C_g/dz^2$) and (ii) $V_g^{dc}$ enhances the tension in the suspended flake \cite{hari}. The first one dominates for our devices. At finite values of $V_g^{dc}$, resonant frequency $f = \sqrt{(f_0)^2 - \alpha (V_g^{dc})^2}$. $\alpha$ depends upon the device geometry but not on $E$.
\bibitem{strain correction}
Cr, Au and NbSe$_2$ \cite{thermal} have positive thermal expansion coefficients. On heating from 28 K to 40 K, they will expand, and the strain $\varepsilon$ will reduce. If any correction due to thermal expansion is incorporated, then the estimate of change in $E$ will be even greater than 10 per cent.
\end{thebibliography}

\newpage

\end{document}